\documentstyle[11pt]{article}
\begin{document}
\setlength{\baselineskip}{15pt}
\title{One solution of the 3D Jacobi identities allows determining \\ an infinity of them}
\author{Benito Hern\'{a}ndez--Bermejo $^1$}
\date{}

\maketitle

\begin{center}
\noindent{\em Universit\'{e} Libre de Bruxelles (ULB), Service de Physique Th\'{e}orique et 
Math\'{e}matique.} \\
\noindent{\em Campus Plaine -- CP 231, Boulevard du Triomphe, B-1050 Bruxelles, Belgium.}
\end{center}

\mbox{}

\mbox{}

\mbox{}

\begin{center} 
{\bf Abstract}
\end{center}
\noindent
It is demonstrated that the knowledge of a single and arbitrary solution of the 
three-dimension\-al Jacobi equations allows determining infinite families of new solutions, 
which are generally and explicitly constructed in what follows. Examples are given.

\mbox{}

\mbox{}

\mbox{}

\mbox{}


\mbox{}

\mbox{}

\mbox{}

\mbox{}

\mbox{}

\mbox{}

\mbox{}

\mbox{}

\mbox{}

\mbox{}

\mbox{}

\mbox{}

\mbox{}

\mbox{}

\mbox{}

\mbox{}

\noindent $^1$ E-mail: bhernand@ulb.ac.be

\pagebreak
\begin{flushleft}
{\bf 1. Introduction}
\end{flushleft}

Poisson structures (see \cite{olv1} and references therein) have an important 
presence in all fields of Mathematical Physics, such as dynamical systems theory 
\cite{7}-\cite{28}, fluid dynamics \cite{hs1}, plasma physics \cite{25}, optics 
\cite{17,dht3}, etc. Describing a given physical system in terms of a Poisson 
structure opens the possibility of obtaining a wide range of information 
which may be in the form of perturbative solutions \cite{cyl1}, invariants 
\cite{byv3,tbym}, nonlinear stability analysis \cite{jmaa,hyct}, bifurcation 
properties and characterization of chaotic behaviour \cite{dht3}, efficient 
numerical integration \cite{mac1} or integrability results \cite{mag1}, 
to cite a few. 

In this work finite-dimensional Poisson systems defined on the Euclidean 
space $I \! \! R^n$ shall be considered, thus excluding Poisson structures on 
other $n$-dimensional manifolds. Mathematically, a finite-dimensional dynamical 
system defined on $I \! \! R^n$ is said to have a Poisson structure if it can be 
written in terms of a set of ODEs of the form: 
\begin{equation}
    \label{nham}
    \dot{x}_i = \sum_{j=1}^n J_{ij} \partial _j H \; , \;\:\; i = 1, \ldots , n, 
\end{equation} 
where $ \partial_j $ means $ \partial / \partial x_j$ and $H(x)$, which is 
usually taken to be a time-independent first integral, plays the role of Hamiltonian function. 
The $J_{ij}(x)$ are the entries of an $n \times n$ matrix ${\cal J}$ 
which may be degenerate in rank ---known as the structure matrix--- and they 
have the property of being solutions of the Jacobi identities: 
\begin{equation}
     \label{jac}
     \sum_{l=1}^n ( J_{li} \partial_l J_{jk} + J_{lj} \partial_l J_{ki} + 
     J_{lk} \partial_l J_{ij} ) = 0 
\end{equation}
In (\ref{jac}), indices $i,j,k$ run from 1 to $n$. The $J_{ij}$ must also verify the 
additional condition of being skew-symmetric:
\begin{equation}
     \label{sksym}
     J_{ij} =  - J_{ji} \;\:\:\:\: \mbox{for all} \:\; i,j
\end{equation}

The possibility of describing a given vector field not explicitly written in the form 
(\ref{nham}) in terms of a Poisson structure is an obvious question of fundamental importance 
in this context, and is still an open subject \cite{7}-\cite{28},\cite{5}-\cite{21}.
Mathematically, this 
problem amounts to giving an algorithm for decomposing (whenever possible) a C$^1$ function 
$f(x): \Omega \subset I \!\! R^n \longrightarrow I \!\! R^n$, where $\Omega$ is 
open, as $f(x)= {\cal J}(x) \cdot \nabla H(x)$, where 
${\cal J}$ is a solution of the combined system (\ref{jac}-\ref{sksym}), and 
$H(x)$ is a C$^1$ real-valued function. This is a nontrivial decomposition to which 
important efforts have been devoted in past years in a variety of approaches. 
The source of the difficulty is obviously twofold: First, a known constant of motion of the 
system able to play the role of the Hamiltonian is needed. And second, it is necessary to find 
a suitable structure matrix for the vector field. Consequently, finding a solution of the 
Jacobi identities (\ref{jac}) complying also with the additional skew-symmetry conditions 
(\ref{sksym}) is unavoidable. This explains, together with the intrinsic mathematical interest 
of the problem, the permanent attention deserved in the literature by the obtainment and 
classification of skew-symmetric solutions of the Jacobi equations 
\cite{7}-\cite{hyg1},\cite{nut1,nut2,pla1,27},\cite{byv1}-\cite{29}. 
Since the Jacobi identities constitute a set of 
nonlinear coupled PDEs, the characterization of $n$-dimensional solutions has 
followed, roughly speaking, a sequence of increasing nonlinearity. In this way we can speak of 
constant structure matrices (of which the symplectic matrix is a particular case), linear (i.e. 
Lie-Poisson) structures \cite{olv1,29}, affine-linear structures \cite{bha1}, and quadratic 
structures \cite{byv2,pla1,byr1,lyx1}, as well as some very general structure matrices which 
may contain functions of arbitrary nonlinearity \cite{byv4}. However, the set of 
solutions of system (\ref{jac}-\ref{sksym}) seems to be still mostly unexplored. 
Perhaps the only exception to this situation is that of three-dimensional (3D in what follows) 
vector fields, which constitute an important case which has been repeatedly considered in the 
literature and is the best understood at present 
\cite{7}-\cite{hyg1},\cite{17,nut1,nut2,27,8,byv1}. In dimension three, the 
strategy for finding suitable skew-symmetric solutions of the Jacobi equations has often been 
problem-dependent. In this sense, we can find recipes based on the use of either convenient 
{\em ansatzs\/} for the solution \cite{gyn1,nut1,nut2,27}, or symmetry considerations 
\cite{7,17}, or 
the knowledge of additional information about the system, such as the existence of a constant 
of motion \cite{hyg1,8,byv1}. Additionally, in the 3D situation it is also possible to 
recast the problem (\ref{jac}-\ref{sksym}) in equivalent forms which may be more suitable for 
the determination of the desired solutions \cite{9}-\cite{hyg1}. Although still incomplete, this is 
certainly a more elaborate state of affairs than the one existing in the general 
$n$-dimensional case. Moreover, it is worth recalling that the 3D scenario is 
particularly relevant for several reasons: 

\begin{enumerate}
\item A large number of 3D systems arising in very diverse fields 
have a Poisson structure \cite{7}-\cite{hyg1},\cite{17}-\cite{nut2},\cite{27,8}. Therefore 
3D Poisson structures are the natural framework for the analysis of such systems. 
\item Dimension three corresponds to the first nontrivial case where a Poisson structure does 
not imply a symplectic structure. In other words, it is the simplest meaningful kind of 
Poisson structures which is not symplectic. 
\item Three is the lowest dimension for which the Jacobi identities are not 
always identically verified (recall that every skew-symmetric $2 \times 2$ matrix is a 
structure matrix). Since the complexity of equations (\ref{jac}-\ref{sksym}) is increasing with 
the dimension $n$, the 3D case is the simplest nontrivial one as well as a 
natural first approach to the full problem of analyzing system (\ref{jac}-\ref{sksym}).
\end{enumerate}

In this work it is demonstrated that the knowledge of a single arbitrary solution of the 3D 
Jacobi equations (\ref{jac}-\ref{sksym}) allows determining explicitly an infinity of 
additional solutions. As we shall see, the reason is that the existence of one known solution 
of the 3D Jacobi equations (\ref{jac}-\ref{sksym}) implies that the problem of determining new 
solutions can be reformulated in terms of finding families of solutions of one 
linear PDE. Notice that in the 3D case, system (\ref{jac}-\ref{sksym}) consists of only 
one independent PDE, and then the number of PDEs is not being reduced along the process.
However, given that the construction of solutions for a linear PDE can, in principle, be 
done from the analysis of its associated characteristic equations, we are actually reducing the 
3D nonlinear PDE problem (\ref{jac}-\ref{sksym}) to the analysis of several simultaneous ODEs, 
namely the characteristics. In this way it shall be demonstrated that the problem of finding 
new infinite families of solutions of the original nonlinear 3D Jacobi equations 
(\ref{jac}-\ref{sksym}) can always be solved provided one solution is already known. Since many 
particular solutions of (\ref{jac}-\ref{sksym}) have already been identified in the 3D case, 
this leads immediately to a method for the explicit determination of solutions, as intended.

The structure of the article is as follows. In Section 2 the results enumerated above, leading 
to the general and explicit determination of new infinite families of solutions of the 3D 
Jacobi equations, are presented. Section 3 contains several examples which illustrate the 
theory. The work is concluded in Section 4 with some final remarks.

\mbox{}

\begin{flushleft}
{\bf 2. Linear problem associated to the 3D Jacobi equations}
\end{flushleft}

For the sake of conciseness, the following notation for the entries of the 3D 
structure matrix shall be used:
\begin{equation}
	\label{uvw}
	u(x) := J_{12}(x) \; , \;\:\; v(x) := J_{31}(x) \; , \;\:\; w(x) := J_{23}(x)
\end{equation}
In the case $n=3$, system (\ref{jac}-\ref{sksym}) actually consists of a single independent 
equation. If we make use of the definition (\ref{uvw}) we can rewrite such equation in the 
form:
\begin{equation}
     \label{jac3df}
     u \partial_1 v - v \partial_1 u + 
     w \partial_2 u - u \partial_2 w + v \partial_3 w - w \partial_3 v = 0 
\end{equation}
The equation corresponding to the 3D version of system (\ref{jac}-\ref{sksym}) shall be 
written in the form (\ref{jac3df}) in the rest of the work.

Now let $\{ u_0 (x) , v_0 (x) , w_0 (x) \}$ be a known solution of (\ref{jac3df}). In what 
follows it shall be assumed that it is a regular and nontrivial solution, i.e. that the rank of 
the Poisson structure represented by $\{ u_0 (x) , v_0 (x) , w_0 (x) \}$ is constant and equal 
to two everywhere in the domain of interest. Then we can look for new solutions according to 
the {\em anstatz \/} $\{ u_0 (x) + \xi (x) , v_0 (x) + \xi (x) , w_0 (x) + \xi (x) \}$, where 
$\xi (x)$ is an arbitrary C$^1$ function to be determined. If we substitute the {\em anstatz \/} 
in (\ref{jac3df}) we see after some algebra that all the nonlinear terms are of the form 
$\xi \partial _i \xi$, $i=1,2,3$, and in fact all such terms do cancel out. Consequently, we 
arrive to the following {\em linear \/} PDE for $\xi$:
\begin{equation}
     \label{linpde}
     (u_0-v_0) \partial_1 \xi + (w_0-u_0) \partial_2 \xi + (v_0-w_0) \partial_3 \xi  = 
	\lambda (x) \xi 
\end{equation}
where
\begin{equation}
     \label{eigen}
     \lambda (x) = 
     \partial_1 (u_0 - v_0) + \partial_2 (w_0 - u_0) + \partial_3 (v_0 - w_0)
\end{equation}

Therefore, the mere knowledge of one solution of (\ref{jac3df}) allows a threefold 
simplification of the problem of finding new solutions:
\begin{enumerate}
\item We can transform a nonlinear problem into a linear problem.
\item We can transform the PDE problem (\ref{jac3df}) into an ODE problem, namely the one 
given by the characteristic equations of (\ref{linpde}-\ref{eigen}).
\item We can reduce the number of unknowns from three to one.
\end{enumerate}

We can now proceed to analyze equation (\ref{linpde}-\ref{eigen}). Three cases must be 
distinguished:

\pagebreak
\noindent{\em (I) Case $\lambda (x)=0$. \/} 

It is relatively frequent, as we shall see in the examples section, that 
$\{ u_0 (x) , v_0 (x) , w_0 (x) \}$  are such that $\lambda (x)$ in (\ref{linpde}-\ref{eigen}) 
vanishes at every point of the domain of interest. In such case the characteristic equations 
of (\ref{linpde}) are:
\begin{equation}
\label{char1}
	\frac{\mbox{d}x_1}{u_0 - v_0}=\frac{\mbox{d}x_2}{w_0 - u_0}=\frac{\mbox{d}x_3}{v_0 - w_0} 
	\:\;\: , \;\: \mbox{d} \xi =0
\end{equation}
Obviously, we need two constants of motion of (\ref{char1}) in order to have the general 
solution of (\ref{linpde}-\ref{eigen}). It is not difficult to verify that two such constants 
can be chosen as $K_1(x)=x_1+x_2+x_3$ and $K_2(x)=C(x)$, where $C(x)$ is a Casimir invariant of 
the known solution $\{ u_0 (x) , v_0 (x) , w_0 (x) \}$. Accordingly, the general solution of 
(\ref{linpde}-\ref{eigen}) is:
\begin{equation}
\label{sol11}
	\xi (x) = \Psi (x_1+x_2+x_3,C(x)) 
\end{equation}
where $\Psi$ is an arbitrary C$^1$ function of its two real arguments. Therefore we have 
arrived to the following new family of solutions of (\ref{jac3df}):
\[
	\{ u(x),v(x),w(x) \} = 
\]
\begin{equation}
\label{sol12}
	\{ u_0(x)+\Psi (K_1(x),K_2(x)),v_0(x)+\Psi (K_1(x),K_2(x)),
	w_0(x)+\Psi (K_1(x),K_2(x)) \}
\end{equation}

We can now consider the second possibility.

\mbox{}

\noindent{\em (II) Case $\lambda (x) \neq 0$, with invertible $K_1(x)$ and $K_2(x)$. \/} 

This time the characteristic equations are:

\begin{equation}
\label{char2}
	\frac{\mbox{d}x_1}{u_0 - v_0}=\frac{\mbox{d}x_2}{w_0 - u_0}=\frac{\mbox{d}x_3}{v_0 - w_0} 
	= \frac{\mbox{d} \xi }{\lambda \xi}
\end{equation}

We need three constants of motion of (\ref{char2}) in order to find the general integral 
of (\ref{linpde}-\ref{eigen}). However, the two constants known from Case I, i.e. 
$K_1(x)=x_1+x_2+x_3$ and $K_2(x)=C(x)$ (where $C(x)$ is a Casimir invariant of the known 
solution $\{ u_0 (x) , v_0 (x) , w_0 (x) \}$) are also first integrals of (\ref{char2}). The 
third constant of motion required is then evident and takes the form of a quadrature: Assuming 
that the standard invertibility conditions arise for $K_1(x)$ and $K_2(x)$, it will be possible 
to make use of both invariants and express two independent variables in terms of the remaining 
one, $K_1$ and $K_2$. For instance:
\begin{equation}
\label{elim}
	x_2 = \alpha(x_1,K_1,K_2) \;\:\:\; , \;\:\: x_3 = \beta(x_1,K_1,K_2)
\end{equation}
Then it is immediate to write:
\begin{equation}
	\frac{\mbox{d} \xi }{ \xi } = \frac{ \lambda \mbox{d} x_1}{u_0-v_0} \equiv 
	\kappa(x_1,K_1,K_2) \mbox{d} x_1
\end{equation}
After integration we finally obtain the third constant of motion:
\begin{equation}
\label{quad3}
	K_3(x) = \frac{ \xi }{ H(x_1,K_1(x),K_2(x))} 
\end{equation}
where
\begin{equation}
\label{fdquad3}
	H(x_1,x_2,x_3)= \exp{\left( \int \kappa(x_1,x_2,x_3) \mbox{d} x_1 \right) }
\end{equation}
Consequently, from (\ref{fdquad3}) we arrive to the following general integral for $\xi$ in the 
case $\lambda \neq 0$:
\begin{equation}
\label{genint}
	\Phi \left(K_1(x),K_2(x),\frac{ \xi }{ H(x_1,K_1(x),K_2(x))} \right) =0
\end{equation}
where $\Phi$ is an arbitrary C$^1$ function of its three real arguments.

We finally have:

\mbox{}

\noindent{\em (III) Case $\lambda (x) \neq 0$, with non-invertible $K_1(x)$ and $K_2(x)$. \/} 

It may still happen that the usual invertibility conditions are not satisfied for $K_1(x)$ and 
$K_2(x)$, i.e. it is not possible to determine equations of the form (\ref{elim}). In this case 
it is still feasible to easily construct infinite families of solutions from a given one. As 
usual, $\{ u_0 (x) , v_0 (x) , w_0 (x) \}$ denote the known solution, corresponding to a 
structure matrix ${\cal J}(x)$. It is well known that after a diffeomorphic change of variables 
$y=y(x)$ a structure matrix ${\cal J}(x)$ is transformed into another structure matrix 
${\cal J'}(y)$ according to the tensor rule:
\begin{equation}
	\label{jdiff}
      J'_{ij}(y) = \sum_{k,l=1}^3 \frac{\partial y_i}{\partial x_k} J_{kl}(x) 
	\frac{\partial y_j}{\partial x_l}
\end{equation}
In principle, the change of variables (\ref{jdiff}) needs not be globally defined on the domain 
of interest for what is to follow. However, for the sake of simplicity the global character of 
the transformation shall be assumed. In the case of a local change of variables the procedure 
described below is not affected, the only difference being that we would arrive to new families 
of solutions of the Jacobi equations defined locally on a subset of the initial domain of 
definition. 

Therefore, a new system of coordinates in which (\ref{linpde}-\ref{eigen}) can be solved for 
${\cal J'}(y)$ is to be introduced. This is very simple to do, but obviously there is not a 
unique choice. For instance, a straightforward possibility is the Darboux canonical form of the 
matrix ${\cal J}(x)$, i.e.
\begin{equation}
\label{darb3d}
	{\cal J'}(y) = \left( \begin{array}{ccc} 0 & 1 & 0 \\ -1 & 0 & 0 \\ 0 & 0 & 0 
				\end{array} \right)
\end{equation}
According to (\ref{uvw}) and (\ref{jdiff}), in the case of the choice (\ref{darb3d}) we are 
mapping $\{ u_0 (x) , v_0 (x) , w_0 (x) \}$ into 
\begin{equation}
	\{ u_0' (y) , v_0' (y) , w_0' (y) \} = \{ J'_{12}(y), J'_{31}(y), J'_{23}(y) \} = 
	\{ 1,0,0 \}
\end{equation}
Notice that equations (\ref{linpde}-\ref{eigen}) become trivial for (\ref{darb3d}) because in 
the Darboux form we are in Case I, actually. Then, the general solution of 
(\ref{linpde}-\ref{eigen}) for matrix (\ref{darb3d}) is $\xi (y) = \Psi (y_1+y_2+y_3,y_3)$. In 
this way we have arrived to the family of solutions:
\[
	\{ u'(y),v'(y),w'(y) \} = \{ u_0'(y),v_0'(y),w_0'(y) \} + 
	\Psi (y_1+y_2+y_3,y_3) \{1,1,1\} =
\]
\begin{equation}
\label{soliii}
	 \{ 1,0,0 \} + \Psi (y_1+y_2+y_3,y_3) \{1,1,1\}
\end{equation}
in evident notation. Then we can make use of (\ref{jdiff}) and transform back the solution 
family $\{ u'(y),v'(y),w'(y) \}$ in (\ref{soliii}) into the original coordinates $x_i$. We 
thus arrive to:
\[
\{ u(x),v(x),w(x) \} = \{ u_0(x),v_0(x),w_0(x) \} + 
\]
\begin{equation}
	\label{soliiix}
	\Psi (y_1(x)+y_2(x)+y_3(x),y_3(x)) \{ M_{12}(x),M_{31}(x),M_{23}(x) \} 
\end{equation}
where
\[
	M_{ij}(x)= \sum_{k,l=1}^3 \frac{\partial x_i}{\partial y_k} A_{kl} 
	\frac{\partial x_j}{\partial y_l} \;\:\; , \;\:\;\: i,j=1,2,3
\]
with the $A_{kl}$ being the entries of
\begin{equation}
\label{aiii}
	{\cal A} = \left( \begin{array}{ccc} 0 & 1 & -1 \\ -1 & 0 & 1 \\ 1 & -1 & 0 
				\end{array} \right)
\end{equation}
This completes the procedure of Case III. Of course, the whole method remains entirely 
identical in the case of choices different to (\ref{darb3d}).

The result just described in Case III is interesting for several reasons:
\begin{itemize}
\item The first one is that we are producing solutions such that the terms added to $u_0$, 
$v_0$ and $w_0$ in (\ref{soliiix}) now are not one and the same due to the presence of the 
functions $M_{ij}$. Clearly this is due to the fact that we are making a previous 
transformation of variables. Therefore the procedure allows determining solutions which are not 
only those of the form $\{ u_0 + \xi , v_0 + \xi , w_0 + \xi \}$, actually. In other words, we 
see that the method is in fact more general that it seemed in principle, leading to more 
general families of solutions than those originally expected. 
\item The second one is that the procedure described in Case III is, in fact, also applicable 
in Cases I and II, because the verification of condition (\ref{elim}) is not relevant 
for the introduction of a new coordinate system. Therefore the generality of the method, as 
considered in the previous item, applies to all cases and is a general feature of this approach 
to the determination of new solutions.
\end{itemize}

We can now proceed to see some examples illustrating the previous possibilities.

\begin{flushleft}
{\bf 3. Examples}
\end{flushleft}

\noindent{\em (I) Constant structure matrices. \/} 

Constant structure matrices are, in spite of their simplicity, ubiquitous in very diverse 
problems, an important example being the Darboux representation of 3D Poisson structures. 
Obviously, every constant 3D skew-symmetric matrix is a structrure matrix. Therefore, let 
$\{ u_0,v_0,w_0 \}$ be constants, not all equal to zero. We then have $\lambda =0$ in 
(\ref{linpde}-\ref{eigen}). Two cases must be distinguished:

\mbox{}

\noindent {\em i) \/} $u_0=v_0=w_0 \neq 0$. In this situation equation (\ref{linpde}-\ref{eigen}) 
becomes trivial and every differentiable $\xi (x)$ is a solution. 

\mbox{}

\noindent {\em ii) \/} $u_0$, $v_0$ and $w_0$ are not equal. This is the generic case. 
According to (\ref{sol11}) we only need to find a Casimir invariant of $\{ u_0,v_0,w_0 \}$. 
It is straightforward to check that one choice is $C(x)=w_0x_1+v_0x_2+u_0x_3$. Consequently 
we arrive to the family of solutions:
\begin{equation}
	\{ u,v,w \} = \{ u_0,v_0,w_0 \} + \Psi (x_1+x_2+x_3,w_0x_1+v_0x_2+u_0x_3) \{1,1,1\}
\end{equation}
in evident notation, with $\Psi$ arbitrary. Notice that $K_1(x)=x_1+x_2+x_3$ and 
$K_2(x)=w_0x_1+v_0x_2+u_0x_3$ are independent when $u_0$, $v_0$ and $w_0$ are not equal, as we 
are now assuming by hypothesis. Thus starting from a trivial, constant solution we have 
found a nonconstant and nontrivial family of solutions just by finding one Casimir invariant. 

\mbox{}

\noindent{\em (II) so(3) and Hamiltonian ray optics structures. \/} 

Another important Poisson structure is the the Lie-Poisson bracket associated to the Lie 
algebra $so(3)$, namely $\{ u_0,v_0,w_0 \}= \{ x_3,x_2,x_1\}$. In this case we again have 
$\lambda =0$, and the resulting PDE (\ref{linpde}-\ref{eigen}) is:
\begin{equation}
     \label{so3pde}
     (x_3-x_2) \partial_1 \xi + (x_1-x_3) \partial_2 \xi + (x_2-x_1) \partial_3 \xi  = 0
\end{equation}
It is well known that a Casimir invariant of this structure is $C(x)=x_1^2+x_2^2+x_3^2$. The 
general solution is then $\xi = \Psi (x_1+x_2+x_3,x_1^2+x_2^2+x_3^2)$. Therefore we have 
found the family:
\begin{equation}
	\{ u,v,w \} = \{ x_3,x_2,x_1 \} + \Psi (x_1+x_2+x_3,x_1^2+x_2^2+x_3^2) \{1,1,1\}
\end{equation}
Consequently the $so(3)$ structure can now be seen as a particular case of a much wider set. 

Many other 3D Poisson structures of the Lie-Poisson kind can also be 
generalized in a very similar way. For instance, we can consider the Hamiltonian ray optics 
structure \cite{17} given by $\{ u_0,v_0,w_0 \}= \{ 4x_3,-2x_1,-2x_2\}$. In this case we again 
have $\lambda =0$. The Casimir invariant is $C(x)=x_1x_2-x_3^2$ and consequently we arrive to 
the family of solutions:
\begin{equation}
	\{ u,v,w \} = \{ 4x_3,-2x_1,-2x_2 \} + \Psi (x_1+x_2+x_3,x_1x_2-x_3^2) \{1,1,1\}
\end{equation}
Therefore the treatment is completely similar to that of $so(3)$, as anticipated.

\mbox{}

\noindent{\em (III) Kermack-McKendric model structure. \/} 

We now take as our starting point the Kermack-McKendric Poisson structure \cite{gyn1,nut2} 
given by $\{ u_0,v_0,w_0 \}= \{ -rx_1x_2,0,-ax_2\}$, with $r$ and $a$ real constants. In 
this case $\lambda$ is not identically zero, and system (\ref{linpde}-\ref{eigen}) becomes:
\begin{equation}
     \label{kmckpde}
      -rx_1x_2 \partial_1 \xi + (rx_1x_2-ax_2) \partial_2 \xi + ax_2 \partial_3 \xi  
	= (r(x_1-x_2)-a) \xi
\end{equation}
Of course we know a first invariant $K_1(x)=x_1+x_2+x_3$, and a second one which is given by 
the Casimir invariant of $\{ u_0 , v_0 , w_0 \}$. This is easily found \cite{byv3,tbym} to be 
$K_2(x)=C(x) = x_3 + (a/r) \ln x_1 $. Therefore we only need the third invariant $K_3$ in 
order to have the general integral of (\ref{kmckpde}). We can first make use of $K_1$ and 
$K_2$ to find relationships of the type (\ref{elim}). In our case, after some algebra we 
obtain:
\begin{equation}
\label{elkmck1}
	x_1 = \exp \left( \frac{r}{a}(K_2-x_3) \right)
\end{equation}
\begin{equation}
\label{elkmck2}
	x_2 = K_1-x_3- \exp \left( \frac{r}{a}(K_2-x_3) \right) 
\end{equation}
Substituting (\ref{elkmck1}-\ref{elkmck2}) in the characteristics we arrive to:
\begin{equation}
\label{sepkmk}
	\frac{\mbox{d} \xi }{ \xi } = \frac{r(x_1-x_2)-a}{ax_2} \mbox{d} x_3 =
	\frac{ rx_3-rK_1-a+2r \exp \left( \frac{r}{a}(K_2-x_3) \right)}{ aK_1-ax_3- 
		a\exp \left( \frac{r}{a}(K_2-x_3) \right)} \mbox{d} x_3
\end{equation}
Integrating (\ref{sepkmk}) and simplifying we can set:
\begin{equation}
	K_3(x) = \frac{\xi }{ x_2 } \exp \left( \frac{rx_3}{a} \right) 
\end{equation}
Therefore the general integral of equation (\ref{kmckpde}) is:
\begin{equation}
\label{igkmck}
	\Phi \left[x_1+x_2+x_3 , x_3 + (a/r) \ln x_1 ,\frac{\xi }{ x_2 } 
	\exp \left( \frac{rx_3}{a} \right) \right] = 0
\end{equation}
with arbitrary differentiable $\Phi$. In spite of its seeming complexity, equation 
(\ref{igkmck}) contains families of very simple solutions. As an example of this assertion, 
it is straightforward to verify that, for instance, the solution $\xi(x)= kx_1x_2$, $k \in 
I \!\! R$, belongs to the general integral (\ref{igkmck}).

\mbox{}

\noindent{\em (IV) Lotka-Volterra and Generalized Lotka-Volterra structures. \/} 

As a final example we consider the Lotka-Volterra \cite{gyn1,nut1,25},\cite{pla1}-\cite{28} 
and Generalized Lotka-Volterra \cite{byv2,jmaa} structures of the form 
\begin{equation}
	\label{uvwglv}
\{ u_0,v_0,w_0 \} = \{ a_{12}x_1x_2,a_{31}x_1x_3,a_{23}x_2x_3 \} 
\end{equation}
where the $a_{ij}$ are real constants for all $i,j$, and $x_i>0$ for all $i$ (i.e. the domain 
of definition of these structures is the interior of the positive orthant of $I\!\!R^3$). For 
the sake of conciseness, we shall consider here the generic case in which none of the $a_{ij}$ 
is zero. It is then easy to check that 
\begin{equation}
	\label{lglv}
	\lambda = (a_{31}-a_{12})x_1 + (a_{12}-a_{23})x_2 +(a_{23}-a_{31})x_3
\end{equation}
does not vanish in general. Therefore we have to make use of the two 
invariants, namely $K_1=x_1+x_2+x_3$ and the Casimir $K_2=x_1^{a_{23}}x_2^{a_{31}}x_3^{a_{12}}$ 
and find two relationships of the form (\ref{elim}). Clearly this is not possible in this case, 
as anticipated in Section 2. Consequently we have to apply the procedure of Case III. 
For this we shall perform a suitable change of variables which is diffeomorphic 
and globally defined in the interior of the positive orthant of $I\!\!R^3$: We define 
$(y_1,y_2,y_3)=(x_1^{\alpha},x_2^{\beta},x_3^{\gamma})$. According to equation (\ref{jdiff})
it is not difficult to show that always there exist suitable values of $\alpha$, $\beta$ and 
$\gamma$ such that $\{ u'_0(y),v'_0(y),w'_0(y) \}$ become either $\{ y_1y_2,y_1y_3,y_2y_3 \}$ 
or $\{ -y_1y_2,-y_1y_3,-y_2y_3 \}$. In both cases, equation (\ref{lglv}) is still applicable in 
the new variables and now we do have $\lambda (y)=0$. Therefore in the variables 
$(y_1,y_2,y_3)$ we are in Case I of Section 2 and thus we are led to the general solution 
$\xi (y) = \Psi (y_1+y_2+y_3,y_1y_2y_3)$, with arbitrary differentiable $\Psi$. If we now 
transform back these results into the original variables $(x_1,x_2,x_3)$ we arrive to the 
general family of solutions given by:
\[
\{ u,v,w \} = \{ a_{12}x_1x_2,a_{31}x_1x_3,a_{23}x_2x_3 \} + 
\]
\begin{equation}
\label{solglv}
	\Psi (x_1^{\alpha}+x_2^{\beta}+x_3^{\gamma},x_1^{\alpha}x_2^{\beta}x_3^{\gamma})
	\{ a_{12}x_1^{1-\alpha}x_2^{1-\beta}, a_{31}x_1^{1-\alpha}x_3^{1-\gamma},
	a_{23}x_2^{1-\beta}x_3^{1-\gamma}\}
\end{equation}

As suggested in Section 2 (Case III) the derivation of solutions taking (\ref{uvwglv}) as 
starting point and making use of the Darboux canonical form (\ref{darb3d}) is another possible 
line of action. Here it has been preferred the use of a different choice in order to emphasize 
the multiplicity of suitable possibilities for the determination of new solutions. However, the 
use of the Darboux canonical form is an equally good and simple alternative in this case. It 
will be omitted here for the sake of brevity, although it is straightforward in the present 
example (see \cite{byv2} for the algorithm of global reduction of the structures (\ref{uvwglv}) 
to the Darboux form). It is worth recalling that the solutions found in the case of the Darboux 
reduction are, of course, different to (\ref{solglv}) since the multiplicity of choices 
reflects the multiplicity of solution families that can be determined.

\mbox{}

\begin{flushleft}
{\bf 4. Final remarks}
\end{flushleft}

As seen in the Introduction, the search of solutions of the Jacobi equations is a current topic 
in the domain of Poisson structures. It has been shown that knowledge of a solution greatly 
simplifies the procedure of determining new solution families in the 3D case. In particular, 
this is possible due to two main reasons:
\begin{itemize}
\item Knowledge of a given solution allows reformulating the problem into a linear one (as it 
can be verified without difficulty, this property only holds in dimension 3). This is 
not the first time that such kind of simplification is presented in the literature regarding 3D 
Poisson systems \cite{hyg1,byv1}. However, relevant advantages of the present method when 
compared to \cite{hyg1,byv1} are:
	\begin{itemize}
	\item Now we operate directly on the Poisson structure independently of the form of the 
	Hamiltonian, while in \cite{hyg1,byv1} an specific Hamiltonian is to be assumed.
	\item The present method produces in a straightforward way a large number of new 
	solution families. In this sense, it seems to be more effective and simpler to apply 
	than \cite{hyg1,byv1}.
	\end{itemize}
\item The second is that the use of a known solution as starting point allows reducing the 
number of unknows from  three to one. Again, this type of reduction in the number of unknowns 
is not new in the field, a good example being the conformal invariance of the solutions of the 
3D Jacobi equations \cite{gyn1}. However, the quantity and richness of solutions produced by 
the method described in the previous sections is very large when compared to the single, 
linearly dependent family of solutions that the conformal invariance generates.
\end{itemize}
Therefore the present approach can be regarded quite naturally in the framework of the 
analysis of the Jacobi identities. However, it is the simultaneous combination of the two 
previous properties in the present method what actually makes it fruitful and simple to 
apply in the determination of new solutions. 

\pagebreak

\begin{flushleft}
{\bf Acknowledgements}
\end{flushleft}

\noindent I would like to acknowledge Prof. V\'{\i}ctor Fair\'{e}n for fruitful discussions. 
This research has been supported by a Marie Curie Fellowship of the European Community 
programme {\em ``Improving Human Research Potential and the Socio-economic Knowledge Base'' \/} 
under contract number HPMFCT-2000-00421.

\pagebreak

\end{document}